\documentclass[twocolumn,showpacs,preprintnumbers,amsmath,amssymb,floatfix,superscriptaddress]{revtex4-1}

\usepackage[dvips]{graphicx}% Include figure files
\usepackage{dcolumn}% Align table columns on decimal point
\usepackage{bm}% bold math
\usepackage{bbm}% bold math
\usepackage{amsmath,amsthm,amssymb}
\usepackage{graphicx}
\usepackage{epstopdf}
\usepackage{psfrag}
\usepackage{color}

\def\beq{\begin{equation}}
\def\eeq{\end{equation}}

\newcommand{\mH}[0]{\mathcal{H}}

\newcommand{\D}[0]{\mbox{\tiny $D$}}

\newcommand{\bp}[0]{\mathbf p}
\newcommand{\bq}[0]{\mathbf q}
\newcommand{\bP}[0]{\mathbf P}
\newcommand{\bb}[0]{\mathbf b}
\newcommand{\be}[0]{\mathbf e}
\newcommand{\bu}[0]{\mathbf u}
\newcommand{\bv}[0]{\mathbf v}
\newcommand{\br}[0]{\mathbf r}
\newcommand{\bR}[0]{\mathbf R}
\newcommand{\bx}[0]{\mathbf x}
\newcommand{\by}[0]{\mathbf y}

\newcommand{\spl}[1]{\begin{align}\begin{split} #1 \end{split} \end{align}}
\newcommand{\al}[1]{\begin{align} #1 \end{align}}
\newcommand{\vvec}[2]{\begin{pmatrix} #1 \\ #2 \end{pmatrix}}
\newcommand{\matr}[4]{\begin{pmatrix} #1 & #2 \\ #3 &  #4 \end{pmatrix}}
\newcommand{\matrnop}[4]{\begin{matrix} #1 & #2 \\ #3 &  #4 \end{matrix}}

\newcommand{\NI}[0]{{N_I}}
\newcommand{\mI}[0]{m_I}
\newcommand{\mA}[0]{m_A}

\newcommand{\NA}[0]{N_{\mbox{\tiny A}}}
\newcommand{\Hph}[0]{H_{\mbox{\tiny ph}}}
\newcommand{\picwide}[3]{
\begin{figure*}[t]
\includegraphics[width=#3]{#1} \caption  {#2}
\end{figure*}
}

\newcommand{\affA}{Institut f{\"u}r Physik, Johannes Gutenberg-Universit{\"a}t Mainz, D-55099 Mainz, Germany}
\newcommand{\affB}{Institut f{\"u}r Quanteninformationsverarbeitung, Universit\"at  Ulm, Albert-Einstein-Allee 11, D-89069 Ulm, Germany }
\newcommand{\affC}{Institut f\"ur Theoretische Physik, Johann Wolfgang Goethe-Universit\"at, 60438 Frankfurt/Main, Germany}
\newcommand{\affD}{Zentrum f\"ur Optische Quantentechnologien, Universit\"at Hamburg, The Hamburg Centre for Ultrafast Imaging, Luruper Chaussee 149, D-22761 Hamburg}
\newcommand{\affE}{Faculty of Physics, University of Warsaw, PL-00-681 Warsaw, Poland}

\begin{document}

\title{ Emulating Solid-State Physics with a Hybrid System of Ultracold Ions and Atoms}

\author{U.~Bissbort}\affiliation{\affC}
\author{D.~Cocks}\affiliation{\affC}
\author{A.~Negretti}\affiliation{\affD}
\author{Z.~Idziaszek}\affiliation{\affE}
\author{T.~Calarco}\affiliation{\affB}
\author{F.~Schmidt-Kaler}\affiliation{\affA}
\author{W.~Hofstetter}\affiliation{\affC}
\author{R.~Gerritsma}\affiliation{\affA}

\email{bissbort@physik.uni-frankfurt.de}

\date{\today}

\begin{abstract}
We propose and theoretically investigate a hybrid system composed of a crystal of trapped ions coupled to a cloud of ultracold fermions. The ions form a periodic lattice and induce a band structure in the atoms. This system combines the advantages of high fidelity operations and detection offered by trapped ion systems with ultracold atomic systems. It also features close analogies to natural solid-state systems, as the atomic degrees of freedom couple to phonons of the ion lattice, thereby emulating a solid-state system.  Starting from the microscopic many-body Hamiltonian, we derive the low energy Hamiltonian including the atomic band structure and give an expression for the atom-phonon coupling. We discuss possible experimental implementations such as a Peierls-like transition into a period-doubled dimerized state.

\end{abstract}


\maketitle

The study of quantum many-body systems through quantum simulators, as initially proposed by Feynman ~\cite{Feynman:1982,Lloyd:1996,Cirac:2012}, is one of the major challenges of experimental quantum physics. Trapped ultracold quantum gases and ions in electromagnetic traps have proven to be among the most successful systems for this task~\cite{Bloch:2012,Blatt:2012}. Both systems are tunable throughout a wide range of parameters and have been used in seminal quantum simulations~\cite{Bloch:2012,Blatt:2012}. Neutral atomic systems, which can exhibit fermionic statistics naturally, have been very successful in analog simulation of solid-state systems. These setups are even able to simulate artificial gauge fields~\cite{Lin:2009b} and control over and detection of individual atoms is now within reach~\cite{Gericke:2008,Bakr:2009,Sherson:2010}. Trapped ionic systems, on the other hand, have exhibited a remarkably precise control of preparation and measurement of arbitrary system states and can be used to construct a universal digital quantum simulator~\cite{Lanyon:2011}. Their scalability, however, remains an important issue and is currently under intense investigation~\cite{Islam:2012, Britton:2012}.

In this Letter we propose a setup to realize an analog quantum simulator, consisting of the combination of a trapped ion crystal interacting with a system of ultracold fermionic atoms [see Fig. 1(a)], combining the advantages of both systems~\cite{Grier:2009,Zipkes:2010,Schmid:2010,Rellergert:2011,Ratschbacher:2012,Ratschbacher:2013}. Specifically, we focus on a 1D ion chain, but higher-dimensional structures could also be used. We envision fermionic atoms simulating electrons in a solid-state system with ions forming the lattice. A crucial component of solid-state systems, namely the phonon-electron coupling, arises naturally in our scenario. This feature is absent in ultracold atoms trapped in optical lattices, where there is no significant backaction of the atoms on the lattice. Likewise, trapped ion systems do not naturally feature fermionic statistics, so that the quantum simulation of coupled fermion-phonon systems would add significant experimental complexity~\cite{Casanova:2012}. The proposed simulator may be used to study models of electron-phonon coupling, such as the Fr\"ohlich model, Peierls transitions and the emergence of phonon-mediated interactions.

\begin{table}[b]
\begin{tabular}{llll}
 & Solid state & $^6$Li-$^{174}$Yb$^+$ & $^{40}$K-$^{40}$Ca$^+$ \\
\noalign{\smallskip}\hline\\
Lattice spacing $d$ (nm) & 0.3--0.6  & 10$^3$--10$^4$ & 10$^3$--10$^4$\\
Length scale $R^*$ (nm) & 0.026 & 71 & 245\\
Energy scale $E^*$ (kHz) & $10^{13}$ & 166 & 2.1 \\
$d/R^*$ & 10--20  & 14--140 & 4--40\\
$m_i/m_f$ & 10$^{4}$--10$^5$ & $29$ & $1.0$\\
Fermi energy (MHz) & 10$^{8}$ & 0.02 & 0.02\\
Phonon energy (MHz) & 10$^{6}$ & 0.01--10 & 0.01--10 \\
\noalign{\smallskip}\hline
\end{tabular}
\caption{Comparison between a typical natural solid-state system and the proposed quantum simulator. The characteristic length and energy scales for atom-ion interaction are defined in the text and correspond to the hydrogen electron for the solid-state system.
The fermionic mass (electron or atom) and ionic core are denoted by $m_f$ and $m_i$, respectively. 
}
\end{table}
Alternative proposals for investigating phonon couplings include self-assembled lattices of ultracold dipolar molecules~\cite{Pupillo:2008} and optically trapped nanoparticles~\cite{Gullans:2012}, whereas backcoupling to cavity photons has also been demonstrated~\cite{Mottl:2012}. Our proposal is complementary to these approaches and offers advantages in the precise control of the couplings and configuration of the system.

\picwide{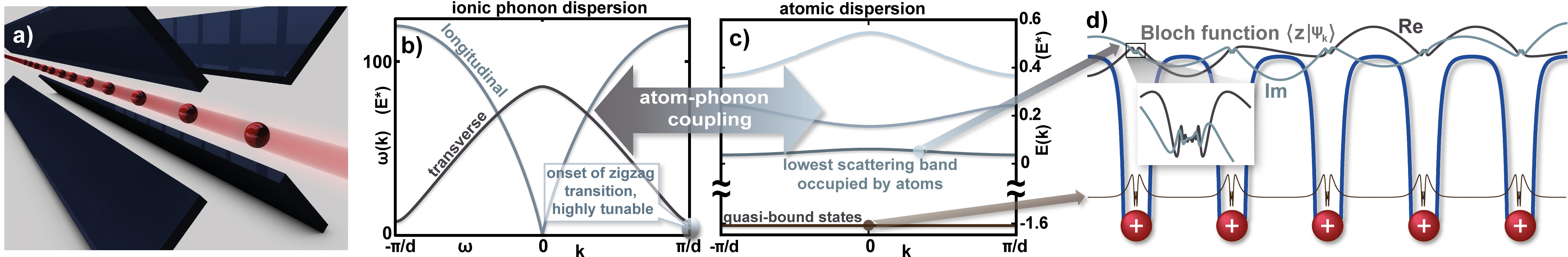}{\label{FIG:setup_and_spectra} (a) A cloud of ultracold atoms is brought close to a linear crystal of ions, which are trapped in a Paul trap. Because of the strong Coulomb repulsion, the ions are strongly pinned to a periodic lattice and the excitations are bosonic phonons, with a typical but highly tunable spectrum shown in (b) in units of $E^*$ for ${}^6$Li -${}^{174}$Yb${}^+$. To lowest order and for transversal atom-ion separation $\Delta_r=0$, mobile atoms experience a periodic spatial potential via the attractive atom-ion potential, forming Bloch states (shown in (d)) and  energy bands (c) for $d=15R^*$ and 1D scattering phases $\varphi_o=-\varphi_e=\pi/4$.\vspace{-5mm}
}{177mm}

The interaction between an atom and an ion is caused by an induced dipole moment in the atom, and for sufficiently large distances is given by $V_{AI}(r) = -\frac{C_4}{r^4}$, where $r$ denotes the atom-ion separation. This defines both a natural length scale $R^*=\sqrt{2 m_A C_4/\hbar^2}$, typically in the 100~nm regime~\cite{Idziaszek:2007}, and an energy scale $E^*=\hbar^4/(4 m_A^2 C_4)$, where $m_A$ is the atomic mass \cite{Note1}.
The interionic Coulomb repulsion leads to phonon branches which also couple to the atoms. Interactions between the atoms are short range and can be tuned, e.g., by Feshbach resonances. In Table~I, we list some typical energy and length scales of our proposed quantum simulator as compared to typical natural solid-state systems. One advantage of our proposal is the flexible mass ratio: a large ratio (e.g. $^6$Li atoms and $^{174}$Yb$^+$ ions) is reminiscent of solid-state systems, whereas equal masses (e.g., $^{40}$K and $^{40}$Ca$^+$) are also possible. 

The outline of this Letter is as follows: we first derive the effective low energy Hamiltonian of the system of spinless ions and atoms, and determine an explicit expression for the atom-phonon coupling. Thereafter we demonstrate the use of our system as a quantum simulator by investigating a Peierls-like transition of a system of fermionic atoms. Finally, we discuss the experimental implementation and further possibilities in an outlook.

\paragraph*{Derivation of the low energy model.} --
The system is microscopically described by three terms in the Hamiltonian $\mH = \mH_I+\mH_A+\mH_{AI}$, where
\spl{
\label{EQ:ion_Hamiltonian}
\mH_I&=\sum_{j=1}^{\NI}\Big[ \frac{\bP_j^2}{2 \mI}+ V_I(\bR_j) \Big] +\frac{e^2}{4 \pi \epsilon_0 } \sum_{\stackrel{i,j=1}{\mbox{\tiny $i>j$}}}^{\NI} \frac{1}{| \bR_i -  \bR_{j}|}
}
describes $\NI$ ions, each of mass $\mI$ and charge $+e$ trapped in a Paul trap represented by the single-particle potential $V_I(\bR_n)$, with pairwise Coulomb interactions. $\bR_n$ and $\bP_n$ are the position and momentum operators of the \mbox{$n$th} ion respectively.
The $\NA$ atoms are described by a Hamiltonian $\mH_A = \sum_{i=1}^{\NA}\Big[ \frac{\bp_i^2}{2\mA}+V_A(\br_i) \Big]$ that includes a tight transverse optical dipole (or magnetic) trap given by $V_A(\br)$ which is oriented parallel to, and can be displaced by a distance $\Delta_r$ from, the center of the Paul trap. In contrast to the case of ions \cite{Note2}, the fermionic quantum statistics of the atoms plays an essential role. The atom-ion interaction is described by the Hamiltonian $\mH_{AI}=\sum_{i,j} V_{AI}(|\br_i-\bR_j|)$.
For the theoretical description, we apply periodic boundary conditions. To derive a low energy effective Hamiltonian, we separately transform the ionic and atomic degrees of freedom into diagonal form and express the coupling between the systems in this basis.

The dominant energy scale is set by the interionic repulsion, which, in conjunction with the trapping potential $V_I$ and a given axial ionic lattice spacing $d$ (typically of the order $1$--$20\mu$m) defines the classical  equilibrium position of the $j$th ion $\overline \bR_j$. This classical configuration minimizing the electrostatic energy $E_I^{(0)}$ corresponds to a linear string, planar zigzag or structures of three-dimensional crystals~\cite{James:1998}.  As the fluctuations of quantum or thermal origin in the ion displacement are small, the ionic Hamiltonian is well suited to an expansion in terms of the position fluctuation operators $\delta \bR_j= \bR_j - \overline \bR_j$.
Hence, we write $\mH_I=E_I^{(0)}+\mH_I^{(2)}+\mH_I^{(3)}+ \ldots$ where the first order terms vanish identically and the low energy spectrum and the dynamics are determined by $\mH_I^{(2)}=\left. \sum_n P_n^2/2m_I + \frac 1 2 \sum_{n,n'} V_{n,n'} \delta R_n \delta R_{n'}\right|_0$, where the combined label $n\equiv(j, D)$ designates the $j$th ion in spatial dimension $D$,
$V_{n,n'}=\partial^2 (V_I+V_{II}) / \partial \delta R_n  \partial \delta R_{n'}$ and $V_{II}$ the interionic potential. After transforming into the basis of phonons with bosonic operators $[\alpha_s,\alpha_s^\dag]=\delta_{s,s'}$, where $s$ is a collective mode label \cite{Note3}, the Hamiltonian in units of $\hbar = 1$ takes on the form (see Appendix~B)
\spl{
\label{EQ:HI_expanded}
\mH_I = E_I^{(0)} + \sum_s \omega_s (\alpha_s^\dag \alpha_s^{\phantom{\dag}} + \frac 1 2)+\frac{\mathcal P^2}{2 \tilde m}+\mathcal O (\alpha^3).
}
The spectrum $\omega_s$ for a particular linear ion string is shown in Fig.~\ref{FIG:setup_and_spectra}(b).

In a similar fashion, an expansion in $\delta \bR_j$ is made for the atom-ion operator $\mH_{AI} = \mH^{(0)}_{AI} + \mH^{(1)}_{AI} + \ldots$, where
\begin{align}
\mH_{AI}^{(0)} \hspace{-1mm} = \hspace{-1mm} \sum_{i,j}\frac{-C_4}{|\mathbf{r}_i-\overline \bR_j|^4},\,\,
\mH_{AI}^{(1)}\hspace{-1mm} = \hspace{-1mm}4C_4 \hspace{-1mm} \sum_{i,j}\frac{(\mathbf{r}_i-\overline \bR_j) \hspace{-0.5mm}\cdot \hspace{-0.5mm} \delta\bR_j}{|\mathbf{r}_i-\overline \bR_j |^6}.
\end{align}
To lowest order $\mH_{AI}^{(0)}$, the atoms feel a periodic lattice of ions pinned to their classical equidistant positions and the higher orders ($\mH_{AI}^{(1)}$, etc.) describe couplings to ionic phonon modes. The atomic eigenstates of $\mH_A + \mH_{AI}^{(0)}$ are Bloch states $\phi_{k,\alpha}(\br)$ with energy $\epsilon_{k,\alpha}$ and characterized by a quasimomentum $k$ and a band index $\alpha$. The form of the long-range atom-ion potential $V_{AI}(r)$ is unbounded from below and the low-lying energy bands [see Fig.~\ref{FIG:setup_and_spectra}(c)] are essentially superpositions of local bound states of each ion with flat dispersions
(bandwidth $\leq 20$~Hz for Yb${}^+$-Li with $d=1$~$\mu$m). Furthermore, they are energetically separated by a large gap [see Fig.~\ref{FIG:setup_and_spectra}(c)] and should not play a significant role if the atoms are moved into the ion string in an appropriate sequence~\cite{Doerk:2010}. On the other hand, the Bloch states with energies above the threshold [see the upper state in Fig.~\ref{FIG:setup_and_spectra}(c)] have a considerably larger bandwidth of typically a few kHz. We expect that adiabatic loading of the atoms into the ion system will populate the lowest of these scattering bands. As it is well separated in energy from other bands, we can work within a single-band approximation and henceforth omit the band index. Alternatively, the system may be prepared into this band by either cooling the atoms or the ions as long as the coupling to the molecular bands is low. When the atom and ion traps are separated by a large distance $\Delta_r$, the atomic band structure is bounded from below and it is most relevant to consider the band with lowest energy.

The single-atom part of the Hamiltonian is $\mH_A + \mH_{AI}^{(0)}=\sum_k \epsilon_k \, c_k^\dag  c_k^{\phantom{\dag }}$, where $c_k^\dag = \int d^3r \, \phi_k(\br) \, \Psi^\dag(\br) $ creates an atom in Bloch mode $k$. As the long-range atom-ion coupling diverges for small separations, an effective description of the short-range coupling is important. It is possible to determine the effective 1D Bloch states and the corresponding dispersion relation within quantum defect theory, as discussed in Appendix~A. For sufficiently strong transverse confinement, any low energy eigenstate will be well approximated by the form $\psi(z,\rho)=\psi_{1D}(z)\, \phi_\perp(\rho)$ in spatial regions between the ions, where the potential felt by the atoms is almost cylindrically symmetric. Here, $\phi_\perp(\rho)$ is the lowest 2D harmonic oscillator state of the optical dipole trap and the trap is oriented in the $\bm{e}_z$-direction. The effect of the ion can then be characterized by the odd and even 1D short-range phases $\varphi_{o}$ and $\varphi_{e}$~\cite{Idziaszek:2007}.  A typical dispersion relation and Bloch state are shown in Figs.~\ref{FIG:setup_and_spectra}(c) and \ref{FIG:setup_and_spectra}(d) respectively. The effective 1D theory is valid only for atom-ion distances that are larger than the transverse wave packet size $l_{\perp}$. For Li and Yb$^+$, and assuming a transverse atomic trap frequency of $\omega_{\perp}=2\pi\times200$~kHz, we find $l_{\perp}\sim R^*$. Hence for these parameters, the 1D description is applicable only for interion distances $d \gg R^*$ and energies $E \ll E^*$. For tighter atomic confinement or heavier atomic species, the description holds over larger parameter regimes. Obtaining the exact form of the atomic wave function closer to the ion would, however, require a 3D description.

\begin{figure}[b]
\includegraphics[width=\columnwidth]{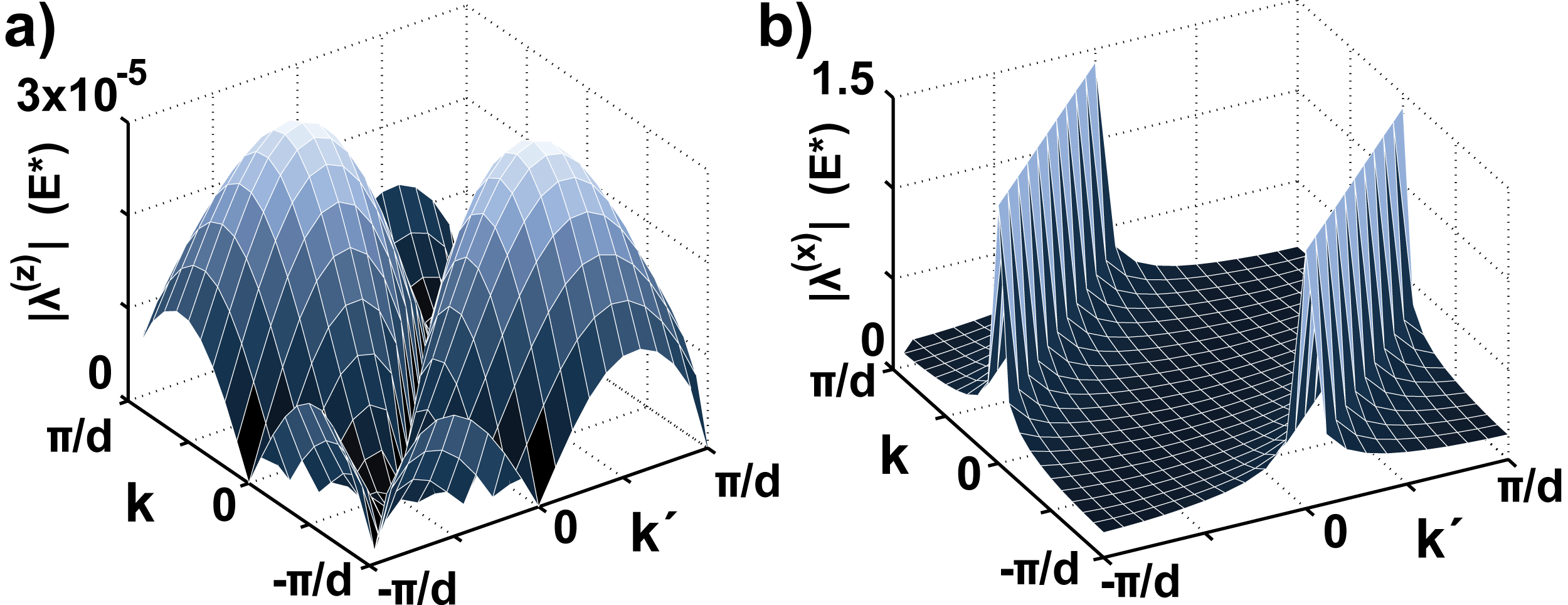} \caption{\label{FIG:coupling}
The interaction between atoms and ions is expressed as an atom-phonon coupling $|\lambda_{k,k'}^{(q=k-k',x/z)}|$ in the low energy limit. For atoms displaced by $\Delta_r=0.75$~$R^*$ and tuning to a small  $\omega_{k=\pi/d,x}=0.0093\Omega_x=30$~$E^*$, the longitudinal coupling (a) is several orders of magnitude smaller than the transverse coupling (b) and the latter may exceed the band width, while still being smaller than the band gap.
}
\end{figure}

\begin{figure*}[t]
\includegraphics[width=180mm]{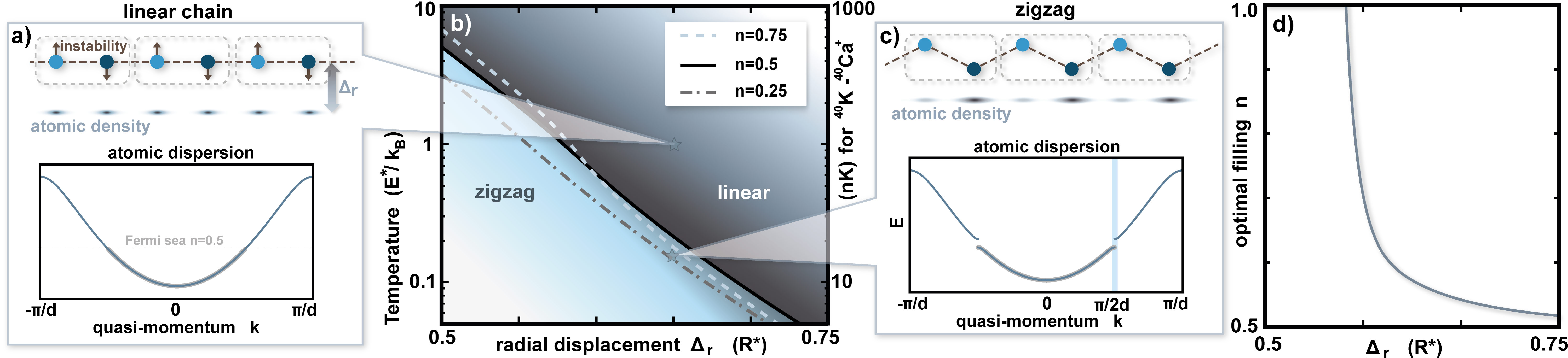} \caption  {\label{FIG:zigzag}Experimentally tuning the ion trapping allows for the observation of a Peierls-like phase transition when atoms are confined in an effective 1D dipole trap at a separation $\Delta_r>0$ from the ion string. The discrete translational symmetry of the linear chain~(a), is spontaneously broken towards a zigzag pattern of the ions~(c) and a band gap opens at $k=\pi/2d$ for the atoms. At different atomic fillings $n$ (per ion), and below a critical temperature $T_c$, a phase transition occurs~(b). Unlike the Peierls effect in solid-state materials, the optimal filling that generates the largest $T_c$ can be larger than $n=1/2$~(d).\vspace{-5mm}}
\end{figure*}

The dominant term of the coupling Hamiltonian is
\spl{
\label{EQ:HAI1}
\mH_{AI}^{(1)}&=\sum_{k,k',s} \frac{\lambda_{k,k'}^{(s)}}{\sqrt{N_I}} \, \alpha_s \,  c_k^\dag \, c_{k'}^{\phantom{\dag}} \; + \; \mbox{H.c.}.
}
It is useful to define the potential $V_n(\br) ={(\br - \overline \bR_{j}) \cdot \be_{D}}/{|\br - \overline \bR_{j}|^6}$, in terms of which the phonon coupling is given by
\begin{equation*}
\label{EQ:atom_phonon_coupling}
\lambda_{k,k'}^{(s)}=4 C_4  \sum_n \frac{v_n^{(s)}-u_n^{(s)}}{\sqrt{2 m_I \Omega_n /N_I} } \int \hspace{-1mm} d^3 \br \, V_n(\br) \, \phi_k^*(\br) \, \phi_{k'}(\br),
\end{equation*}
where $u_n^{(s)}$ and $v_n^{(s)}$ are the phonon mode coefficients and $\Omega_n$  is the bare angular frequency (see Appendix~B). Note that $\lambda_{k,k'}^{(s)}$ is only uniquely defined up to the (arbitrary) complex phases of the individual Bloch and phonon modes.

To calculate $\lambda_{k,k'}^{(s)}$ in the case of overlapping atom and ion traps, we require an accurate description of the Bloch states very close to the ions, which is unfortunately not available in the quantum defect calculation. Furthermore, the secular approximation for the ions neglecting ion micromotion may break down~\cite{Nguyen:2012,Cetina:2012}. We note that a full quantum mechanical treatment of the effect of the time-dependent trapping potential will require a many-body quantum Floquet formalism~\cite{Landa:2012} and is beyond the scope of the present Letter. 

For an atomic cloud tightly confined in the transverse direction and transversally displaced from the ion string by $\Delta_r \be_x$, the low energy single-particle Bloch states are in the transverse ground state and of the form $\psi_{k_z}(\vec{r}) = \psi_{k_z}(z) \phi_\perp(\rho)$.  The 1D Bloch $\phi_k(z)$ states in the potential $V_{1D}(z) = \sum_i V_{AI}(|z \bm{e}_z - \overline{\bR}_i|)$ can be calculated by diagonalizing the Hamiltonian in the plane wave basis. The atom-phonon coupling $\lambda_{k,k'}^{(s)}$, which vanishes unless $q=k-k'$, where $q$ is the quasimomentum corresponding to phonon index $s$ (see Appendix~C), is shown in Fig.~\ref{FIG:coupling}. The transversal coupling features a pronounced peak at $q=|k-k'|=\pi/d$ as the system approaches the zigzag transition, discussed in the following.

\paragraph*{Transverse Peierls-like transition.} --
As a demonstration of our quantum simulator, we investigate a Peierls-type transition \cite{Peierls:1991} at $\Delta_r>0$. It is energetically favorable for a 1D metallic crystal to undergo a transition to an insulating phase, the so-called Peierls transition, when cooled below a critical temperature, as a gap opens in the electronic band structure. For particular fermionic filling, this decrease in energy is enough to offset the increase due to the ion distortion. In our case, the energy that the linear ion string gains as it is distorted into a period-doubled zigzag configuration is offset by the lowering of the energy of the atomic system at half-filling [see Fig.~\ref{FIG:zigzag}(c)]. The structural phase transition from a linear to a zigzag configuration in trapped ions has previously been the subject of theoretical study, as it can be coupled to quantized fields~\cite{Shimshoni:2011,Cormick:2012}. The present proposal goes one step further by using a {\it fermionic} quantum field to induce the phase transition, thereby emulating the Peierls effect as observed in natural solid-state systems.

Fixing a separation $\Delta_r$ between the atom and ion traps, we treat the ion positions classically. The ions are perturbed from their equilibrium positions by an amount $\bm{\delta}$ oriented along the trap separation such that $\overline{\bR}_i^\prime \equiv \overline{\bR}_i + (-1)^n \bm{\delta}$, giving rise to an ionic energy increase $\Delta E_I = m_I \omega_{k=\pi/d,x}^2 \delta^2 / 2 + O(\delta^4)$, with $\delta = |\bm{\delta}|$, associated with a zigzag mode frequency $\omega_{k=\pi/d,x}$. For $T \rightarrow 0$, zigzag order will always be preferred as the band gap scales linearly with $\delta$ and the energy of classical ion displacement increases with $\delta^2$, for small $\delta$. At finite temperature, the preferred configuration is determined by minimizing the free energy, calculated for the combined ion and noninteracting atomic system by parametrically varying $\delta$.

A sample band structure for noninteracting ${}^{40}$K atoms near a $^{40}$Ca$^+$ ion string is shown in Fig.~\ref{FIG:zigzag}(c) where a zigzag displacement $\delta \approx 0.019R^*$ minimizes the energy $U = \langle \mH_A + \mH_{AI}^{(0)} + E_I^{(0)} \rangle$ of a system with atomic half-filling and static ion positions $\overline{\bR}_i^\prime$ with $\omega_{k=\pi/d,x} \approx 2\pi\times60$~kHz, ion spacing $d=2$~$\mu$m~$\approx8 R^*$ and atomic temperature $T_A=15$nK$\approx 0.15E^* / k_B$. The phase diagram for noninteracting $^{40}$K with $^{40}$Ca$^+$ is shown in Fig.~\ref{FIG:zigzag}(b) for varying temperature and atom-ion trap separation. Furthermore, while the largest critical temperatures occur for half-filling when $\Delta_r$ is large, the optimal critical temperature, shown in Fig.~\ref{FIG:zigzag}(d), occurs for larger fillings per ion $n$ as $\Delta_r \rightarrow 0$.

\paragraph*{Implementation.} --
For the proposed quantum simulation, the  interacting quantum systems can be prepared individually and their mutual coupling strength can be controlled by varying the distance $\Delta_r$ between ions and fermions. The ion crystal can be cooled to its ground state using resolved sideband cooling. The motional state of the ion crystal can be initialized and read out by laser-induced coupling to internal states of the ions which can be measured with standard fluorescence measurements~\cite{Leibfried:1996}. Building on these techniques, interferometric methods have been proposed to probe ion crystals near structural phase transitions~\cite{Chiara:2008}.

Precise control over small numbers of ultracold fermions in optical traps has recently been demonstrated~\cite{Serwane:2011}. The atomic density profile is accessible via absorption imaging in combination with time-of-flight analysis. For low densities, the atoms can be retrapped in an optical molasses trap to prevent them from leaving the imaging region during detection. While we have focused on fermionic atomic species in this Letter, it is also possible to trap bosonic species such as $^7$Li, where ion-controlled tunneling dynamics and Josephson physics may be investigated~\cite{Gerritsma:2012}.

Hybrid ultracold atom-ion systems pose several challenges for experimental realization. Inelastic collisions -- in particular charge transfer and molecular ion formation -- may limit the lifetime. It has been observed that for some species the rates may be sufficiently small to study intricate setups, such as in a recent experiment where the spin relaxation of a single Yb$^+$ ion immersed in a bosonic Rb bath has been studied~\cite{Ratschbacher:2013}. However, no experimental data are available for inelastic rates between ultracold K or Li with trapped ions. The use of polarized fermionic systems may reduce the effects of three-body recombination as binary atom collisions are suppressed in the ultracold regime~\cite{Harter:2012}. In the geometry in which we studied the Peierls transition, ions and atoms are separated and inelastic collisions are suppressed. For the reduction of atomic heating effects due to ion micromotion we rely on the largest experimentally feasible mass ratio between ions and atoms ($^{174}$Yb$^+$ and $^6$Li)~\cite{Cetina:2012} and prevent the atoms from approaching the ions too closely~\cite{Nguyen:2012}. To further mitigate micromotion-induced heating, it may be feasible to use heavier molecular ions, which could be sympathetically cooled by suitable cotrapped ion species. 

In conclusion, we have proposed an analog quantum simulator for lattice Hamiltonians that can be tuned into a regime of strong coupling to vibrational phonons. This allows for the simulation of complex many-body Hamiltonians that even go beyond seminal condensed matter models, such as the Fr\"ohlich and Holstein models, known to be relevant for phenomena such as superconductivity.
Future work will focus on solving the 3D many-body atom-ion problem including short-range interactions, considering the time-dependent potential of the ions and the role of the spin degrees of freedom in order to engineer entanglement between the atoms and ions.

This work was supported by the DFG via Sonderforschungsbereiche SFB-TR/49 and SFB-TR/21, Forschergruppe FOR 801, the excellence cluster ``The Hamburg
Centre for Ultrafast Imaging'' of the DFG, and by the EU Projects PICC and SIQS. R.G. acknowledges fruitful discussions with S.T.~Dawkins and K.~Jachymski.

%%-------------------------------

%-------------------------------

\clearpage
\newpage

\begin{appendix}

\section{1D band structure within quantum defect theory}\label{ap_QDT}
Here, the short-range behavior of the atom-ion interaction is described by quantum defect parameters and we work in the 1D approximation~\cite{Idziaszek:2007}. For this to apply, all energy scales in the system have to be small compared to the radial oscillator frequencies, such that the probability for any atom to be in a radially excited state is strongly suppressed.  In this situation the Schr\"odinger equation in units of $R^*$ and $E^*$  for a single atom interacting with an ion located at $\bar{\mathbf{R}}=0$ for $z \gg l_{\perp}=\sqrt{\hbar/(2m_A\omega_{x,y})}$ is given by
\begin{equation}\label{eq_schr_1d}
\left(-\frac{d^2}{dz^2}-\frac{1}{z^4}\right)\psi (z)=E\psi (z).
\end{equation}

\noindent Here we assumed that the ionic mass is much larger than the atomic mass, $m_I \gg m_A$, such that $m_A \approx \mu$, the reduced mass. For  a given finite energy  $E$,  this can be neglected in the limit of small atom-ion separation $z\rightarrow 0$,  and the two-dimensional solution space approaches the asymptotic solutions
\begin{eqnarray}\label{eq_asymp_a}
\tilde{\psi}_e(z)&=&|z|\sin \left(\frac{1}{|z|}+\varphi_e\right)\\
\label{eq_asymp_b}
\tilde{\psi}_o(z)&=&z \sin \left(\frac{1}{|z|}+\varphi_o\right),
\end{eqnarray}
which are characterized by the even and odd phases  $\varphi_{e}$ and $\varphi_{o}$. These effective 1D phase shifts fully characterize the short-range interaction within the 1D regime and have a non-trivial dependence on the 3D s-wave scattering length. Within quantum defect theory, an energy eigenfunction is described by a superposition of  the asymptotic solutions  Eq.~(\ref{eq_asymp_a}) and (\ref{eq_asymp_b}) in the vicinity of each ion up to a sufficiently small distance $r_\epsilon \gg l_{\perp}$.  Outside $r_\epsilon$, the two solutions $\psi_e(z)$ and $\psi_o(z)$  are propagated numerically up to the border of the real space unit cell for a fixed set of $(E,\varphi_e,\varphi_o)$  with the initial conditions chosen to match the asymptotic solutions smoothly at $r_\epsilon$.  The two symmetric and anti-symmetric solutions span the corresponding solution space, guaranteeing that the Bloch state can be written as
\spl{
\psi_k(z)=c_{e}^{(k)}\,{\psi}_e(z) + c_{o}^{(k)}\,{\psi}_o(z).
}
with coefficients $c_{e}^{(k)}$ and $c_{o}^{(k)}$ for the even and odd states respectively.
Furthermore, in accordance with the Bloch theorem, each Bloch state fulfills
\spl{
\psi_k(z+d)=e^{ikd} \, \psi_k(z)
}
under spatial translation by a lattice distance $d$. If $E$ does not lie within a band gap, the associated quasi-momentum $k$ is defined by the constraint of both the wave function and its derivative being continuous at the edge of the unit cell. If the ion is chosen to lie in the middle of the unit cell at $z=0$, symmetry implies $\psi_e(-\frac d 2)=\psi_e(\frac d 2)$, $\psi_o(-\frac d 2)=-\psi_o(\frac d 2)$, $\psi_e'(-\frac d 2)=-\psi_e'(\frac d 2)$, $\psi_o'(-\frac d 2)=\psi_o'(\frac d 2)$ and the matching conditions imply the linear relations

\spl{
A(k) \vvec{c_e^{(k)}}{c_o^{(k)}} = \vvec{0}{0},
}
where
\spl{
A(k) = \left( \rule{0cm}{2em} \matrnop{(1-e^{ikd}) \psi_e(\frac d 2)}{(1+e^{ikd}) \psi_o(\frac d 2)\:}{(1+e^{ikd}) \psi_e'(\frac d 2)}{(1-e^{ikd}) \psi_o'(\frac d 2)} \right).
}
For normalizable, non-trivial solutions of (${c_e^{(k)}},{c_o^{(k)}}$) to exist, the determinant of the coefficient matrix $A(k)$ has to vanish. Thus, the quasi-momentum corresponding to the chosen energy $E$ is efficiently determined using a root finding algorithm on the determinant $\det(A(k))$. The Bloch states on the entire lattice are subsequently given by combining them piecewise in every real-space unit cell.

For a single atom-ion system, the energy eigenstates can be classified into a set of discrete bound states (decaying exponentially in real-space for large distances) and a continuum of unbound scattering states. In the limit of large intra-ionic spacings $d\gg R^*$ the Bloch states  converge to these bound and scattering states and the band index takes on the role of the local eigenstate label. Only the bands corresponding to scattering states would have a significant band width (or correspondingly, significant  hopping elements within the Wannier representation) and the Bloch functions are exponentially suppressed in the regions between the ions. With decreasing $d$ (or fixed $d$ and decreasing atomic mass $m_A$) the classification into bound and scattering states loses its applicability and also the Bloch states emerging from the  bound states obtain a finite band width.

\section{The ionic phonon transformation}
\label{App:phonon_transformation}
Here we discuss the generalized unitary transformation~\cite{{Blaizot:1986}} into the basis of phonon modes in both the linear ion chain, as well as the zigzag regime.
In the low energy regime the higher order terms  are small and the low energy excitations are determined by the collective mode structure of $\mH_I^{(2)}$. Instead of obtaining the eigenmodes in the classical limit, it is favorable to perform the transformation to the phonon basis on an operator level, which yields the precise transformation relations and avoids the problems of ordering of higher order terms before quantizing the bosonic fields. Recalling that the matrix elements are $V_{n,n'} = \partial^2 V_I/\partial \delta R_n \partial \delta R_{n'}$, and defining the local harmonic oscillator frequencies $\Omega_n=\sqrt{V_{n,n}/m_I}$ and the associated annihilation operator for each ion and spatial dimension $b_n=\sqrt{{\mI \Omega_n}/{2}} [\delta X_n + i P_{n} /  ({\mI \Omega_n} )  ]$ (here $\hbar=1$), the ion Hamiltonian can be written as
\spl{
\label{EQ:H_I_quadratic_in_bs}
\mH_I = E_0+ \frac 1 2 {\vvec{\bb}{\bb^\dag}}^\dag    \Hph  \vvec{\bb}{\bb^\dag} + \mathcal O (b^3).
}
Here, $\bb$ ($\bb^\dag$) denote the column vectors of all annihilation (creation) operators respectively and the phonon coefficient matrix is of the form $\Hph=\matr{h}{\Delta }{{\Delta}^*}{{h}^*}$ with the $3N_I$-dimensional submatrices $\Delta_{n,n'}=\frac{ (1-\delta_{n,n'}) V_{n,n'}}{2 \mI \Omega_n}$ and $h_{n,n'}=\delta_{n,n'} \, \Omega_n + \Delta_{n,n'}$. The phonon spectrum and operators are obtained by a symplectic diagonalization, which preserves the bosonic commutation relations of the phonon operators and reduces to a canonical transformation in the classical limit. If $\bx^{(s)}=\vvec{\bu^{(s)}}{-\bv^{(s)}}$ is an eigenvector obtained from the diagonalization to the positive phonon energy $\omega_s$, the associated phonon annihilation operator is defined by $\alpha_s=\bx^{(s)^\dag} \Sigma \vvec{\bb}{\bb^\dag}$. Here, $\Sigma=\matr{\mathbbm 1}{0}{0}{-\mathbbm 1}$ defines the metric for this symplectic (bosonic) space. The elements $u_n^{(s)}$ and $v_n^{(s)}$, appearing in Eq.~(5), are then the $n$-th elements of the subvectors $\bu^{(s)}$ and $\bv^{(s)}$ respectively. However, only $3N_I-1$ phonon modes exist, i.e. the matrix $\Sigma \Hph$ only possesses $3N_I-1$ linearly independent eigenvectors. The missing eigenvector is fundamentally related to the existence of a Goldstone mode and can only occur if there is one zero eigenvalue. Let $\bp$ be the eigenvector of $\Hph$ to an eigenvalue zero. Its norm vanishes with respect to the norm defined by $\Sigma$ and we normalize it with respect to the euclidean norm $\bp^\dag \bp=1$. To span the entire vector space, we additionally define the vector $\bq$, which also has a vanishing $\Sigma$-norm $\bq^\dag \Sigma \bq=0$, is orthogonal to all eigenvectors $\bx^{(s)}$ and $\by^{(s)}$ with respect to $\Sigma$ and furthermore fulfills $\bq^\dag \Sigma \bp=i$. The scalar products
\al{
\mathcal P&=\bp^\dag \Sigma \vvec{\bb}{\bb^\dag}\\
\mathcal Q&=-\bq^\dag \Sigma \vvec{\bb}{\bb^\dag}
}
defines a generalized collective momentum and position operator respectively. Furthermore the relation $\bq^\dag \Sigma \bp=\frac{1}{\tilde m}$ defines a generalized mass $\tilde m$, appearing in the Hamiltonian in Eq.~(2). Since the eigenvectors and $\bq$ span the entire space, the phonon operators $\{\alpha_s\}$, $\{\alpha_s^\dag \}$ together with $\mathcal P$ and $\mathcal Q$ span the same operator space as the original $\{b_n\}$ and $\{b_n^\dag\}$ operators. Thus any operator acting on the Hilbert space of the ions can be expressed exactly in terms of the former and in this sense Eq.~(2) is the exact expression of the full many-body ion Hamiltonian to second order, requiring the term $\mathcal P^2/(2\tilde m)$.

We now explicitly consider the phonon structure and spectrum for the linear ion chain in the limit of a large system. Here, the second order Hamiltonian can be expressed in terms of the creation and annihilation operators $a_{k,\D}=N_I^{-1/2} \sum_j e^{-ikdj} b_{l,\D}$ and their conjugates as
\spl{
\label{EQ:H_I_quadratic_in_a}
\mH_I = E_0+ \frac 1 2 \sum_{\D=x,y,z}\sum_{k} {\vvec{a_{-k,\D}}{a_{k,\D}^\dag}}^\dag    H^{k,\D} {\vvec{a_{-k,\D}}{a_{k,\D}^\dag}} + \mathcal O (a^3).
}
The $2\times 2$ coupling matrices for the longitudinal and transversal case are given by
\spl{
H^{k,z}&=\matr{\Omega_z-2 f_{k,z} }{-2 f_{k,z}}{-2 f_{k,z}}{\Omega_z-2 f_{k,z}}\\
H^{k,x/y}&=\matr{\Omega_z+f_{k,x} }{f_{k,x}}{f_{k,x}}{\Omega_z+ f_{k,x}}
}
respectively. Here we defined
\spl{
f_{k,\D}=\frac{ e^2}{ 4   \pi \epsilon_0  m_I \Omega_{\D} d^3} \mbox{Re}\Big[ \mbox{Li}_3(e^{-i kd }) \Big],
}
where $\zeta(s)$ is the Riemann zeta function, $\mbox{Li}_s(x)$ is the polylogarithm.

The eigenmode energies are found to be
\spl{
\omega_{l}(k)&=\frac{e}{\sqrt{\pi \epsilon_0 m_I d^3}}  \sqrt{\zeta(3)-\mbox{Re}\big( \mbox{Li}_3(e^{-ikd})\big)}\\
\label{omega_transverse}
\omega_{t}(k)&=\sqrt{\omega_I^2 -   \frac{e^2}{2\pi \epsilon_0 m_I d^3}  \Big[  \zeta(3) - \mbox{Re}\big(\mbox{Li}_3(e^{-ikd})\big) \Big]}
}
for the longitudinal and transverse phonon branches respectively, where $\omega_I$ is the transverse ion trapping frequency.
If the two-dimensional eigenvector of the matrix $\Sigma H^{k,\D}$ is denoted by $\vvec{u_{k,D}}{-v_{k,D}}$, the explicit form of the phonon operators is given by 
\spl{
\alpha_{k,\D}^\dag=u_{k,\D} \, a_{k,\D}^\dag + v_{k,\D} \, a_{-k,\D}.
}

For an isolated ion system, the instability towards the formation of zigzag pattern (i.e. a macroscopic occupation of the $\pi/d$ transverse phonon mode) appears if the expression in the root of Eq.~(\ref{omega_transverse}) becomes negative, i.e. approximately for
\spl{
\omega_I< \sqrt{\frac{2.1036 \, e^2}{2 \pi \epsilon_0 m_I d^3}}.
}

In an implementation we have to consider finite ion crystals. As an example we could have 30 Ca$^+$ ions trapped in a Paul trap with radial trapping frequencies of $2\pi\times6$~MHz. The zig-zag mode can then be tuned to the low frequencies considered in the manuscript by accurately setting the axial trapping frequency around $2\pi\times 0.5$~MHz. The ions in the center of the trap will have near-uniform spacing of about 2.1~$\mu$m. Although this setup is quite challenging, it may be tuned further by introducing additional trapping electrodes to engineer potentials that result much more closely in equidistant ion spacing~\cite{Lin:2009}. For Yb$^+$ ions in a trap with the same trapping frequencies, inter-ion distances of 1.3~$\mu$m are reached.

\section{Calculation of the atom-phonon coupling for displaced atoms}
\label{APP:coupling}
For the case of the atom cloud's radial confinement being much smaller than the displacement $\Delta_r$, the integrand is only significantly non-zero on the tube where the atoms are located (enforced by the radial density $|\phi_\perp(r)|^2$ appearing as a factor) and the effective potential becomes a function of $z$ only: $V_n(\br)=V_n(z)$. For the longitudinal phonon mode $D(n)=z$ this is of the explicit form
\spl{
V_j^{(z)}(z)=\frac{z-d j }{[(z-d j)^2+\Delta_r^2]^{3}}.
}
Without loss of generality, we have assumed that the separations between the traps is in the $x$-direction and for $D(n)=x$ we have
\spl{
V_j^{(x)}(z)=\frac{\Delta_r}{[(z-d j)^2+\Delta_r^2]^{3}}
}
and otherwise $V_j^{(y)}=0$.
Here we define the function $F^D_q(z)=\sum_j e^{iq dj } V_j^{(D)}(z)$, where $D=x,z$ for the transverse and longitudinal modes respectively, as well as the periodic function $\tilde F^D_q(z) = e^{-iqz} \, F^D_q(z)$, which fulfills $\tilde F^D_q(z+d)=\tilde F^D_q(z)$. Expressing each Bloch state as $\phi_k(z)=e^{ikz} u_k(z)$, where $u_k(z)$ is also a periodic function $u_k(z)=u_k(z+d)$, the phonon coupling becomes
\spl{
\label{EQ:lambda_qz}
\lambda_{k,k'}^{(q,D)}&=\frac{4 C_4 (v_{q,D} - u_{q,D})}{\sqrt{2 m_I \Omega_D }}  \delta_{BZ(k-k'-q),0} \\
&\times N_I\int_0^d  dz \, F^D_q(z)  \, \phi_k^*(z) \, \phi_{k'}(z) ,
}
where the function $BZ(p)$ folds any momentum vector $p$ back into the first Brillouin zone by adding an integer multiple of the reciprocal lattice vector $2\pi/d$. The Kronecker delta reflects the overall quasi-momentum conservation of the combined atom-ion system. Since the Bloch states are normalized with respect to integration over the entire lattice consisting of $N_I$ sites, the second line of Eq.~(\ref{EQ:lambda_qz}) rapidly converges to a value independent of $N_I$, as the lattice size is increased.
Note that whereas a Bloch state $\phi_k(z)$ is independent of the Brillouin zone $k$ is in $\phi_k(z)=\phi_{k+2\pi/d}(z)$, this is not the case for the function $u_k(z)$.
In principle, an additional term $\sum_{k,k'} \lambda_{k,k'}^{(\mathcal Q)} \mathcal Q c_k^\dag \, c_{k'}^{\phantom{\dag}}$ appears in Eq.~(4), which corresponds to a coupling of the atomic system to the center of mass motion of the ions. However, a calculation shows that the coefficient $\lambda_{k,k'}^{(\mathcal Q)}$ vanishes and hence we exclude it from the low energy Hamiltonian.

\end{appendix}

\end{document}